\documentclass[aps,prl,twocolumn,superscriptaddress,showpacs,groupedaddress]{revtex4}
\usepackage[T1]{fontenc}
\usepackage[latin9]{inputenc}
\usepackage{verbatim}
\usepackage{textcomp}
\usepackage{amsthm}
\usepackage{graphicx}
\usepackage{amsfonts}
\usepackage{amssymb}
\usepackage{times,txfonts}
\usepackage{hyperref}
\usepackage{color}

\usepackage{CJK}

\begin{document}
\begin{CJK*}{UTF8}{gbsn}

\title{Revealing Chern number from quantum metric}

\author{Anwei Zhang}
\email{zawcuhk@gmail.com}
\affiliation{Department of Physics, The Chinese University of Hong Kong, Shatin, New Territories, Hong Kong, China}
\affiliation{Department of Physics, Ajou University, Suwon 16499, Korea}

\begin{abstract}
Chern number is usually characterized by Berry curvature. Here, by investigating the Dirac model of even-dimensional 
 Chern insulator,
we give the general relation between Berry curvature and quantum metric, which indicates that the Chern number can be encoded in quantum metric as well as the surface area of the Brillouin zone on
 the hypersphere embedded in  Euclidean parameter space. 
We find that there is  a corresponding relationship  between the quantum metric and the metric on such a hypersphere.  We show the geometrical property of quantum metric.
 Besides, we give a protocol to measure the quantum metric in the degenerate system. 
 
\end{abstract}


\maketitle
\end{CJK*}\narrowtext


 \noindent\emph{\bfseries Introduction.} Quantum mechanics and geometry are closely linked. When a quantum system evolves adiabatically along a cycle in parameter space, its quantum state will acquire a measurable phase that depends only on the shape of the cycle. This phase is the geometric phase \cite{b1,b2} which is one of the most important concepts in quantum mechanics and the basis of a variety of phenomena and applications. The geometric phase is actually the holonomy in fiber bundle theory \cite{ho1} and when the quantum system evolves under an infinitesimal adiabatic cycle, it is proportional to the symplectic form on the projective space of normalized quantum states, i.e., Berry curvature. Berry curvature is a central quantity in characterizing the topological nature of quantum matter. For instance, the topological invariant: the first  \cite{fi} and second \cite{se} Chen number, are respectively defined by  integrating  the Berry curvature and its wedge product over a closed manifold in parameter space.

Berry curvature corresponds to the imaginary part of a more general quantity: quantum geometric tensor. It was introduced in order to equip the projective space of normalized quantum states with a distance \cite{dis}. 
This effort results in the quantum (Fubini-Study) metric tensor which is the real part of quantum geometric tensor and describes the U($1$) gauge invariant quantum distance between neighboring quantum states.
The notion of geometric phase, Berry curvature, and quantum metric tensor can be generalized to the quantum system with degenerate energy spectra which has a far richer structure. For a given energy level, the wave function is multi-valued around the cycle in the parameter space. As a  result, the geometric phase becomes non-Abelian holonomies which depend on the order of consecutive paths \cite{hol} and Berry curvature becomes a matrix-valued vector. Besides, the adiabatic evolution of the quantum degenerate system will lead to the non-Abelian quantum metric tensor \cite{ma} which measures a local U($n$) gauge invariant quantum distance between two neighboring quantum states in parameterized Hilbert space.

Quantum metric tensor plays an important role in the recent studies of quantum phase transitions \cite{pt1,pt2} and quantum information theory \cite{inf1,inf2}. For example, the diagonal components of the  quantum metric tensor
are actually fidelity susceptibilities, whose critical behaviors are of great interest \cite{int1,int2}. Quantum metric tensor also appears in the research of Josephson junctions \cite{cn},  non-adiabatic quantum evolution \cite{na}, Dirac and tensor monopoles  in Weyl-type systems \cite{gold}, and bulk photovoltaic effect in topological semimetals \cite{bpe,bpe2}. Recently, it was shown that for two-dimensional topological insulator, the quantum metric is related with Berry curvature \cite{r21,r11,r31,pozo}.  For a higher-dimensional topological insulator, the system will be degenerate. Whether the quantum metric is linked with Berry curvature is still unknown. Besides, up to now, the role of quantum metric in the research of topological invariant for higher-dimensional topological insulator has not been explored.

In this paper, by investigating the quantum metric in Dirac model of  2$N$-dimensional  Chern insulator, we find that  the quantum metric is equivalent to the metric on  the hypersphere embedded in Euclidean parameter space. 
 We show that there is a general relation between Berry curvature and quantum metric, which reveals that the Chern number is linked with quantum metric as well as the surface area of Brillouin zone on the hypersphere. We also show the geometrical property of quantum metric.
 Furthermore, we give the scheme to extract the quantum metric in our degenerate system.



\begin{figure}[t]
\centering
\includegraphics[width=0.37\textwidth ]{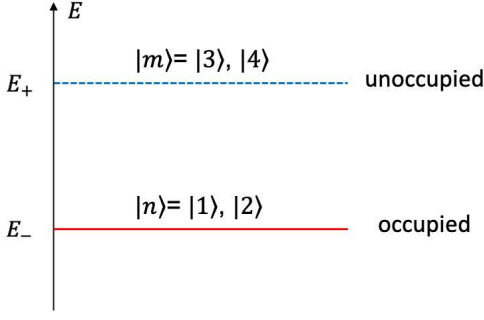}
\caption{Schematic illustration for the energy spectrum with a pair of doubly degenerate bands.
}\label{figure.1}
\end{figure}

\smallskip
 \noindent\emph{\bfseries Corresponding between quantum metric and the metric on sphere.}
We start with  considering the Dirac model of four-dimensional 
  insulator.
This model is widely used. It is a minimal model for four-dimensional topological insulator \cite{qi} and  appears in Dirac Hamiltonian, nuclear quadrupole resonance, and four-band Luttinger model... etc.
This Dirac Hamiltonian has the form
\begin{equation}\label{1}
H(\textbf{k})=d_0(\textbf{k})+\sum^{5}_{i=1}d_i(\textbf{k}) \Gamma_i,
\end{equation}
where $d_0(\textbf{k})$ and $d_i(\textbf{k})$ are real functions of the parameter (momentum) $\textbf{k}=(k_1,k_2,k_3,k_4)$, and $\Gamma_i$ are four-by-four Dirac matrices which satisfy the anti-commutation relations $\{\Gamma_i, \Gamma_j\}=2\delta_{ij}$.
We assume this system has time reversal symmetry, then Kramers theorem implies that the energy bands have two-fold degeneracy with energies $E_{\pm}=d_0(\textbf{k})\pm\sqrt{\sum^{5}_{i=1}d^2_i(\textbf{k}) }$. The  eigenstates of each energy are labeled by $|n\rangle=|1\rangle, |2\rangle$ and $ |m\rangle=|3\rangle, |4\rangle$, respectively (see Fig.~\ref{figure.1}).  
Here we suppose that only the lower pair of the bands are occupied. 

The geometric properties of the occupied degenerate eigenstates $|n_1\rangle$ and $|n_2\rangle$ can be described by the non-Abelian quantum geometric tensor which is defined by \cite{ma}
\begin{equation}\label{2}
Q^{n_1n_2}_{ab}=\sum_{m\in unocc}\langle \partial_a n_1|m\rangle\langle m|\partial_b n_2\rangle,
\end{equation}
where the derivative $\partial_a$ is taken with respect to the parameter $k_a$. The non-Abelian quantum geometric tensor can be decomposed into symmetric and antisymmetric parts: $Q^{n_1n_2}_{ab}=g^{n_1n_2}_{ab}-\frac{i}{2} F^{n_1n_2}_{ab}$, where
  $g^{n_1n_2}_{ab}=\frac{1}{2}(Q^{n_1n_2}_{ab}+Q^{n_1n_2}_{ba})$
is the non-Abelian quantum metric measuring the distance between nearby quantum states,
  $F^{n_1n_2}_{ab}=\partial_a A^{n_1n_2}_b-\partial_b A^{n_1n_2}_a-i[A_a, A_b]^{n_1n_2}$ corresponds to the non-Abelian Berry curvature, and $A^{n_1n_2}_a=i\langle n_1|\partial_a|n_2\rangle$ is the Berry connection of the occupied states $|n_1\rangle$ and $|n_2\rangle$.

\begin{figure}[t]
\centering
\includegraphics[width=0.341\textwidth ]{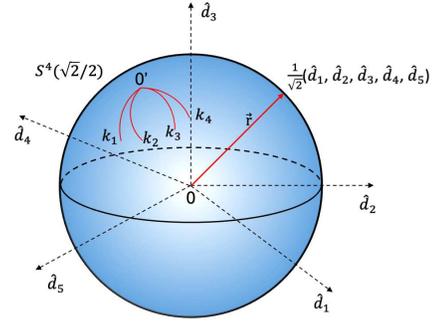}
\caption{Four-dimensional hypersphere embedded in five-dimensional Euclidean parameter space.  $\frac{1}{\sqrt{2}}(\hat{d}_1, \hat{d}_2, \hat{d}_3, \hat{d}_4, \hat{d}_5)=\vec{r}$  and $(k_1,k_2,k_3,k_4)$ are respectively the Cartesian coordinate  and the Gauss coordinate of each point on the hypersphere  with radius $\sqrt{2}/2$. The metric is $\partial_a \vec{r} \cdot \partial_b \vec{r}$ which measures the distance between two neighbouring points on the hypersphere.
}\label{figure.2}
\end{figure}

Now we give the quantum metric in our model. For $m\neq n$, we have the relation:
$\langle n|\partial_a m\rangle(E_m-E_n)=\langle n|\partial_a H|m\rangle$. Substituting this relation into Eq.~(\ref{2}), one obtains
\begin{equation}\label{4}
  Q_{ab} \equiv \sum_{n\in occ}Q^{nn}_{ab} =\frac{1}{4d^2}\sum_{n\in occ}\sum_{m\in unocc}\langle n|\partial_a H|m\rangle\langle m| \partial_b H| n\rangle,
\end{equation}
where $d=(E_+-E_-)/2=\sqrt{\sum^{5}_{i=1}d^2_i(\textbf{k})}$. Then by using $\sum_{m\in unocc}|m\rangle\langle m|=1-\sum_{n'\in occ}|n'\rangle\langle n'|$, we can rewrite Eq.~(\ref{4}) as
\begin{equation}\label{5}
  Q_{ab}=\frac{1}{4d^2}\sum_{n\in occ}\big[\langle n|\partial_a H\partial_b H|n\rangle-\langle n|\partial_a H|n\rangle\langle n|\partial_b H|n\rangle\big].
\end{equation}
Here we have used the relation $\langle n|\partial_a H|n'\rangle=\langle n|\partial_a H|n\rangle \delta_{nn'}$ which results form the fact that $E_{n'}=E_{n}$. Next, we inset Eq.~(\ref{5}) into the definition of quantum metric
\begin{equation}
  g_{ab} \equiv \sum_{n\in occ}g^{nn}_{ab} =\frac{1}{2}(Q_{ab}+Q_{ba})
\end{equation}
 and use the Hellmann-Feynman formula $\langle n|\partial_a H|n\rangle=\partial_a E_n$, this yields
\begin{eqnarray}\label{6}
  g_{ab}&=&\frac{\sum^{5}_{i=1}\partial_a d_i(\textbf{k})\partial_b d_i(\textbf{k})-\partial_a d \partial_b d}{2d^{2}}\nonumber\\
  &=&\frac{1}{2}\sum^{5}_{i=1}\partial_a \hat{d}_i\partial_b \hat{d}_i,
\end{eqnarray}
where $\hat{d}_i=d_i(\textbf{k})/d$. This quantum metric is independent of the overall energy shift $d_0(\textbf{k})$ and 
can be regarded as the metric on a four-dimensional hypersphere embedded in a Euclidean parameter  space (see Fig.~\ref{figure.2}), due to the fact that $\sum^{5}_{i=1}(\hat{d}_i/\sqrt{2})^{2}=(\sqrt{2}/2)^{2}$.

The above results can be generalized to  $2N$-dimensional topological insulator system  which is defined by
\begin{equation}
H(\textbf{k})=\sum^{2N+1}_{i=1}d_i(\textbf{k}) \Gamma_i.
\end{equation}
Here $\textbf{k}$ is $2N$-dimensional parameter and $\Gamma_i$ are $2^N\times 2^N$ matrices satisfying $\{\Gamma_i, \Gamma_j\}=2\delta_{ij}$. The eigenvalues of the Hamiltonian are $\pm\sqrt{\sum^{2N+1}_{i=1}d^2_i(\textbf{k}) }$.
Following similar method, it can be found that the quantum metric for the $2N$-dimensional topological insulator system is 
\begin{equation}
g_{ab}=2^{N-3}\sum^{2N+1}_{i=1}\partial_a \hat{d}_i\partial_b \hat{d}_i,
\end{equation}
where $\hat{d}_i=d_i(\textbf{k})/\sqrt{\sum^{2N+1}_{i=1}d^2_i(\textbf{k}) }$.
This quantum metric correspondes to the metric on the $2N$-dimensional hypersphere embedded in ($2N+1$)-dimensional $\hat{d}_i$ space. Besides, the radius of the hypersphere is $2^{(N-3)/2}$ which increases with the dimension of the 
system.

It is worth noting that for the two-band system, i.e., $N=1$, $\Gamma_i=\sigma_i$ is a Pauli matrix, and energy bands are non-degenerate.  Thus, we should use Abelian quantum geometric $ Q_{ab}=\sum_{m\in unocc}\langle \partial_a n|m\rangle\langle m|\partial_b n\rangle$ and corresponding  Abelian quantum metric  in the derivation of quantum metric.

\noindent\emph{\bfseries The link between quantum metric and Chern number.}  
We now consider the determinant of the quantum metric and its relation with the Chern number.  We first investigate 
 four-dimensional case.
Without changing the determinant, we extend the four-by-four  quantum metric $2g_{ab}$ to a 
five-by-five matrix 
\begin{equation}  
\left(
\begin{array}{ccccc}\label{a}
 1 & 0 & 0 & 0 & 0 \\
0 &\partial_1 \hat{d}_i\partial_1 \hat{d}_i &\partial_1 \hat{d}_i\partial_2 \hat{d}_i & \partial_1\hat{d}_i\partial_3 \hat{d}_i &  \partial_1\hat{d}_i\partial_4 \hat{d}_i \\
0&  \partial_2\hat{d}_i\partial_1 \hat{d}_i  & \partial_2 \hat{d}_i\partial_2 \hat{d}_i  &\partial_2 \hat{d}_i\partial_3 \hat{d}_i &\partial_2 \hat{d}_i\partial_4 \hat{d}_i  \\
0 & \partial_3 \hat{d}_i\partial_1 \hat{d}_i  &\partial_3 \hat{d}_i\partial_2 \hat{d}_i & \partial_3 \hat{d}_i\partial_3\hat{d}_i & \partial_3 \hat{d}_i\partial_4 \hat{d}_i  \\
0 & \partial_4 \hat{d}_i\partial_1 \hat{d}_i & \partial_4 \hat{d}_i\partial_2 \hat{d}_i  &\partial_4\hat{d}_i\partial_3\hat{d}_i  & \partial_4 \hat{d}_i\partial_4 \hat{d}_i 
\end{array}
\right),
\end{equation}
where Einstein summation convention is used for the index $i$. Such a matrix can be decomposed into the product of a matrix $A$ and its transpose matrix $A^{\mathrm{T}}$
\begin{equation}  
\left(
\begin{array}{ccccc}
 \hat{d}_ 1& \hat{d}_ 2& \hat{d}_ 3& \hat{d}_4& \hat{d}_5\\
\partial_1 \hat{d}_1&\partial_1 \hat{d}_2 & \partial_1 \hat{d}_3&\partial_1 \hat{d}_4 & \partial_1 \hat{d}_5 \\
\partial_2 \hat{d}_1& \partial_2 \hat{d}_2&  \partial_2 \hat{d}_3&  \partial_2 \hat{d}_4&  \partial_2 \hat{d}_5\\
\partial_3 \hat{d}_ 1&  \partial_3 \hat{d}_2 &  \partial_3 \hat{d}_3 &  \partial_3 \hat{d}_4 &  \partial_3 \hat{d}_5 \\
\partial_4 \hat{d}_1&  \partial_4 \hat{d}_2 & \partial_4 \hat{d}_3 &\partial_4 \hat{d}_4  & \partial_4 \hat{d}_5\\
\end{array}
\right)
\left(
\begin{array}{ccccc}
 \hat{d}_ 1 & \partial_1 \hat{d}_1& \partial_2 \hat{d}_1 &\partial_3 \hat{d}_ 1 & \partial_4 \hat{d}_1 \\
\hat{d}_ 2 & \partial_1 \hat{d}_2& \partial_2 \hat{d}_2& \partial_3 \hat{d}_ 2 &\partial_4 \hat{d}_2\\
\hat{d}_ 3& \partial_1 \hat{d}_3 & \partial_2 \hat{d}_3 & \partial_3 \hat{d}_ 3& \partial_4 \hat{d}_3  \\
\hat{d}_ 4 & \partial_1 \hat{d}_4  & \partial_2 \hat{d}_4 & \partial_3 \hat{d}_ 4&\partial_4 \hat{d}_4 \\
\hat{d}_ 5 &\partial_1 \hat{d}_5& \partial_2 \hat{d}_5& \partial_3 \hat{d}_5 & \partial_4 \hat{d}_5
\end{array}
\right).
\end{equation}
The determinant of the matrix $A$ is $\epsilon_{\alpha\beta\gamma\eta\tau} \hat{d}_{\alpha}\partial_1 \hat{d}_{\beta}\partial_2 \hat{d}_{\gamma}\partial_3 \hat{d}_{\eta}\partial_4 \hat{d}_{\tau}$,
while the determinant of the matrix in Eq.~(\ref{a}) is $2^4 $ det $g$, thus one has
\begin{eqnarray}\label{7}
  \sqrt {\mathrm{det} ~g}&=&\frac{1}{4}\vert\epsilon_{\alpha\beta\gamma\eta\tau} \hat{d}_{\alpha}\partial_1 \hat{d}_{\beta}\partial_2 \hat{d}_{\gamma}\partial_3 \hat{d}_{\eta}\partial_4 \hat{d}_{\tau}\vert\nonumber\\
  &=&\frac{1}{48}\vert \epsilon_{abcd}\mathrm{tr}(F_{ab}F_{cd})\vert.
\end{eqnarray}
The Ref.~\cite{zh} is used in the last setp. 
If the sign of the above quantities in the  symbol of absolute value  is knowable, for instance, the sign of the quantity $\epsilon_{abcd}\mathrm{tr}(F_{ab}F_{cd})$, i.e., $\mathrm{sgn}(TF)$,  
we can have the second Chern number
\begin{equation}\label{98}
 C_2
 =\frac{1}{S^4}\int_{BZ}  \mathrm{sgn}(TF)~dS,
\end{equation}  
where $S^4=2\pi^2/3$ is the surface area of the four-dimensional hypersphere in Fig.~\ref{figure.2} and  $dS= \sqrt {\mathrm{det} ~g}~d^4k$  is the area measure on the hypersphere. 

If the function $\mathrm{sgn}(TF)$ in Eq.~(\ref{98}) does not depend on  the parameter $\textbf{k}$,  Eq.~(\ref{98}) can be simplified to 
\begin{equation}\label{981}
 C_2
 =\frac{S_{BZ}}{S^4} \mathrm{sgn}(TF),
\end{equation}  
where $S_{BZ}=\int_{BZ}\sqrt {\mathrm{det} ~g}~d^4k$ denotes the surface area of Brillouin zone on the hypersphere.  When $S_{BZ}$ covers the hypersphere $n$ times, Eq.~(\ref{981}) becomes $n$ $\mathrm{sgn}(TF)$ which provides a new way for us to calculate the Chern number  in addition to the traditional method, such as the method in Ref.  \cite{cpb}. On the other hand, if the function $\mathrm{sgn}(TF)$ dependes on  the parameter $\textbf{k}$, Eq.~(\ref{98})  will be
\begin{equation}\label{982}
 C_2
 =\frac{S_{BZ+}-S_{BZ-}}{S^4},
\end{equation}  
where $S_{BZ\pm}$ denote the surface area of Brillouin zone  with $\mathrm{sgn}(TF)=\pm1$, respectively.

In the same way, we can have the determinant of the quantum metric in $2N$-dimensional topological insulator system
\begin{eqnarray}\label{91}
  \sqrt {\mathrm{det} ~g}&=&2^{N(N-3)}\vert\epsilon_{\alpha_1\alpha_2\ldots \alpha_{2N+1}} \hat{d}_{\alpha_1}\partial_1 \hat{d}_{\alpha_2}\partial_2 \hat{d}_{\alpha_3}\ldots\partial_{2N} \hat{d}_{ \alpha_{2N+1} }\vert \nonumber\\
  &=&\frac{2^{N^2-3N+1}}{(2N)!}\vert \epsilon_{a_1a_2\ldots a_{2N}}\mathrm{tr}(F_{a_1a_2}\ldots F_{a_{2N-1}a_{2N}})\vert
\end{eqnarray}
and the $N$th Chern number
\begin{equation}\label{92}
 C_N=\frac{1}{S^{2N}}\int_{BZ} \mathrm{sgn}(NF)~ dS,
\end{equation}  
where $S^{2N}=\pi^NN!2^{N^2-N+1}/(2N)!$ is the surface area of the $2N$-dimensional hypersphere, $\mathrm{sgn}(NF)$ denotes the sign of the quantity $\epsilon_{a_1a_2\ldots a_{2N}}\mathrm{tr}(F_{a_1a_2}\ldots F_{a_{2N-1}a_{2N}})$ and $dS$
refers to area measure $\sqrt {\mathrm{det} ~g}~d^{2N}k $ on the $2N$-dimensional hypersphere.  This result reveals that the Chern number is connected to quantum metric as well as the surface area of Brillouin zone on the hypersphere.
For two-dimensional system, i.e., $N=1$, Eq.~(\ref{91}) is reduced to
\begin{equation}\label{92}
 \sqrt {\mathrm{det} ~g}=\frac{1}{2}\vert F_{12}\vert,
\end{equation}  
which is agreement with the result in Ref. \cite{r21,r11,r31,pozo}.

\noindent\emph{\bfseries Geometrical property of quantum metric.} 
Next we continue to explore the properties of the quantum metric.  We focus on four-dimensional case.
With the quantum metric in Eq.~(\ref{6}), the Levi-Civita connection $\Gamma^a_{bc}=\frac{1}{2}g^{ad}(\partial_b g_{dc}+\partial_c g_{db}-\partial_d g_{bc})$, curvature tensor $R^a_{bcd}=\partial_c \Gamma^a_{bd}-
\partial_d \Gamma^a_{bc}+ \Gamma^a_{ec} \Gamma^e_{bd}- \Gamma^a_{ed} \Gamma^e_{bc}$  and covariant curvature tensor $R_{abcd}=g_{ae}R^e_{bcd}$ in parameter space can be constructed. This covariant curvature tensor is actually the covariant curvature tensor that defines the curvature of the four-dimensional hypersphere in Fig.~\ref{figure.2}. For embedded  hypersphere  in Euclidean parameter space,  according to the Gauss-Codazzi equations \cite{gauss},  the covariant curvature tensor has the form
\begin{equation}\label{10}
 R_{abcd}=2(g_{ac}g_{bd}-g_{ad}g_{bc}),
\end{equation}
where the reciprocal of the coefficient $2$ is the square of the radius.
Since the dimension of the quantum metric is four,  from Eq.~(\ref{10}) we have the Ricci curvature tensor $R_{ab}=6g_{ab}$,  the Ricci scalar $R=24$ and the equation
\begin{equation}\label{11}
 R_{ab}-\frac{1}{2}g_{ab}R+6g_{ab}=0,
 \end{equation} 
which has the similar form with vacuum Einstein equation.
Here the quantum metric is positive-definite \cite{ooo} and the hypersphere in Fig.~\ref{figure.2} is a sub-manifold in Euclidean parameter space rather than real space. 
We note that the gravitational instanton \cite{gra} in quantum theories of gravity is the solution of vacuum Einstein equation with positive-definite metric.

The Euler characteristic number $\chi$ measures the topological nature of manifold. In the four-dimensional case, it is proportional to $\int_{BZ}d^{4}k \frac{1}{\sqrt {\mathrm{det} ~g}}\epsilon_{klmn}\epsilon_{abcd}R_{klab}R_{mncd}$ \cite{euler}. From Eq.~(\ref{10}), one can derive $\epsilon_{klmn}\epsilon_{abcd}R_{klab}R_{mncd}=384\mathrm{det} ~g$. Thus the Euler characteristic number  $\chi$ is proportional to the area  of Brillouin zone on the four-dimensional hypersphere $S_{BZ}=\int_{BZ}\sqrt {\mathrm{det} ~g}~d^4k$.

\noindent\emph{\bfseries Implementation.} 
The Hamiltonian Eq.~(\ref{1}) can be realized by using the four hyperfine ground states of rubidium-87 atoms coupled with radio-frequency and microwave fields \cite{r1}. 
Electric circuits  \cite{e1,e2,e3} can also be used to implement this Hamiltonian. Now we consider how to extract the quantum metric in our degenerate system.  Inspired by the measurement scheme for non-degenerate system \cite{non},
we modulate the parameter $k_a$ in time as
\begin{equation}\label{12}
 k_a(t)=k_a+(2\epsilon /\hbar\omega)\cos(\omega t).
 \end{equation} 
According to Fermi's golden rule, the transition rate from lower bands to upper bands is
$\Gamma(\omega)=\frac{2\pi}{\hbar^2}\sum_{m,n}|\frac{\epsilon}{\hbar\omega}\langle m|\partial_a H |n\rangle |^2\delta (\omega_{mn}-\omega)$ and
the integrated
rate $\int d\omega \Gamma(\omega)$ is 
\begin{eqnarray}\label{13}
\Gamma^{\mathrm{int}}&=&\frac{2\pi \epsilon^2}{\hbar^2}\sum_{m,n}\langle \partial_a n|m\rangle\langle m|\partial_a n\rangle\nonumber\\
&=&\frac{2\pi \epsilon^2}{\hbar^2}Q_{aa}=\frac{2\pi \epsilon^2}{\hbar^2}g_{aa}.
 \end{eqnarray} 
In the last step, we have used $\sum_{n}F^{nn}_{ab}=i(Q_{ab}-Q_{ba})=0$ which is derived from Eq.~(\ref{5}).

Similar to the Ref.~\cite{non}, the off-diagonal components of the quantum metric tensor can be
extracted by modulating the two sets of parameters
\begin{eqnarray}\label{14}
 k_a(t)&=&k_a+(2\epsilon /\hbar\omega)\cos(\omega t),\nonumber\\
  k_b(t)&=&k_b\pm(2\epsilon /\hbar\omega)\cos(\omega t)
 \end{eqnarray}
and measuring the differential integrated rate 
\begin{equation}\label{15}
\Delta\Gamma^{\mathrm{int}}=\Gamma^{\mathrm{int}}_{+}-\Gamma^{\mathrm{int}}_{-}=\frac{8\pi \epsilon^2}{\hbar^2}g_{ab},
 \end{equation} 
 where $\Gamma^{\mathrm{int}}_{\pm}=\frac{2\pi \epsilon^2}{\hbar^2}(g_{aa}\pm 2g_{ab}+g_{bb})$ resulting from the perturbing Hamiltonian $(2\epsilon /\hbar\omega)\cos(\omega t)(\partial_a H\pm \partial_b H)$.
Thus, by observing the excited rate of quantum system under a proper time-periodic modulation, we can obtain every component of quantum metric tensor. Such a protocol can be applied to the general even-dimensional topological insulator system.

\noindent\emph{\bfseries Concluding remarks.}
Our work shows the metric - quantum metric duality and reveals the deep connection between quantum metric and Berry curvature  as well as Chern number (and winding number), which is meaningful for exploring the topological nature of quantum matter and studying the band geometry  in topological insulator. However, our results are made for the Dirac model of even-dimensional Chern insulator. For odd-dimensional case, following the similar method, the relation between quantum metric and topological invariant  winding number can also be found easily. Since Dirac model has only two bands with degeneracy or not, thus for multi-band system which will be not described by Dirac Hamiltonian, our results are no longer valid. The relation between quantum metric and Berry curvature for multi-band system is  beyond the scope of the present paper. However,  for two-dimensional multi-band system, it has been known that $\sqrt {\mathrm{det} ~g}$ is no less than $\frac{1}{2}\vert F_{12}\vert$ \cite{roy} which may provide some hints for us to explore the general relation between quantum metric and Berry curvature for multi-band system.

\begin{acknowledgments}
We would like to thank R.-B. Liu for useful discussion and N. Goldman for comment. After reviewing the draft manuscript of the paper, B. Mera and N. Goldman would like to re-derive and generalize it to odd dimensional case in a mathematical way  \cite{bm}.
\end{acknowledgments}


\begin{thebibliography}{81}
\bibitem{ooo} In the hyperspherical coordinates $\hat{d}_ 1=\cos \theta_1$, $\hat{d}_ 2=\sin\theta_1\cos \theta_2$, $\hat{d}_ 3=\sin\theta_1\sin\theta_2\cos \theta_3$, 
 $\hat{d}_ 4=\sin\theta_1\sin\theta_2\sin\theta_3\cos\theta_4$, $\hat{d}_ 5=\sin\theta_1\sin\theta_2\sin\theta_3\sin\theta_4$,  the non-zero components of the quantum metric tensor are
$g_{\theta_1\theta_1}=\frac{1}{2},  g_{\theta_2\theta_2}=\frac{1}{2}\sin^2\theta_1,  g_{\theta_3\theta_3}=\frac{1}{2}\sin^2\theta_1\sin^2\theta_2,
 g_{\theta_4\theta_4}=\frac{1}{2}\sin^2\theta_1\sin^2\theta_2\sin^2\theta_3$. It can be seen that these diagonal elements are positive.
\end{thebibliography}

\end{document}